\documentstyle[aps,prd]{revtex}
\def\bi{\bibitem}
\newcommand{\be}{\begin{equation}}
\newcommand{\ee}{\end{equation}}
\newcommand{\beq}{\begin{eqnarray}}
\newcommand{\eeq}{\end{eqnarray}}
\newcommand{\bear}{\begin{array}}
\newcommand{\ear}{\end{array}}

\begin{document}
\title{Wormhole Dominance Proposal And Wave Function Discord}    
\author{
S.Biswas$^{*a),b)}$, B.Modak$^{a)}$ and A.Shaw$^{**a)}$\\ 
a) Dept. of Physics, University of Kalyani, West Bengal,\\
India, Pin.-741235 \\
b) IUCAA, Post bag 4, Ganeshkhind, Pune 411 007, India \\
$*$  email: sbiswas@klyuniv.ernet.in\\
$**$ email: amita@klyuniv.ernet.in}
\date{}
\maketitle
\begin{abstract}
Using the wormhole dominance proposal, it is shown that quantum corrections to 
the usual WKB ansatz for the wave function of the universe ably circumvent many 
of the drawbacks present in the current proposals. We also find that the recent 
criticism by Hawking and Turok does not apply to the tunneling proposal.
\end{abstract}
\medskip
\bigskip
\par
PACS Number(s): 98.80.Hw, 98.80.Bp
\section{\bf{Introduction}}
There are currently three proposals, namely the no boundary proposal \cite{har:prd}, 
the tunneling proposal \cite{vil:prd} and Linde's proposal \cite{lin:nuo} on the wave 
function of the universe. Recently a fourth proposal \cite{bis:prd} is given and we call                                                    
it wormhole dominance proposal. Apart from the debate that the above three proposals 
do or not lead to sufficient inflation, recently the dispute arises whether the 
pair production of blackholes during inflation would lead to catastrophic 
instability to the de Sitter space or not. It is argued by Bousso and Hawking 
\cite{bou:prd} that the tunneling wave function would lead to catastrophic 
instability to the de Sitter space. This claim was analyzed in ref. \cite{gar:prd} 
and was shown to be unfounded. On the otherhand Linde \cite{lin:gr} 
demands an unacceptably low values for the density parameter if one uses Hartle-
Hawking wavefunction. Hawking and Turok \cite{haw:gr} misunderstood the 
Linde's wavefunction as the tunneling wavefunction and assert that the tunneling wavefunction would lead to 
unstable perturbations about a homogeneous cosmological background and is 
meaningful for homogeneous minisuperspace models.  
\par
Vilenkin \cite{gar:prd} objects to the pair production of blackholes as a 
process of independent nucleation and considers the pair production of black 
holes as pair production of massive particles in de Sitter space. He starts with 
the tunneling wavefunction and obtains the solution of the Wheeler-DeWitt 
equation in the region $m >> H$, where the quantum state of the scalar field 
(which are produced ) are obtained using the tunneling boundary conditions. 
He shows the emergence of de Sitter invariant Bunch-Davies vacuum. Using Mellor 
and Moss' result \cite{mel:pl} for the nucleation rate of blackholes, 
it is shown that the tunneling boundary condition also reproduces this result. 
This makes him conclude that the tunneling wavefunction is not an opposite 
prescription to the Hartle-Hawking wavefunction and since the blackholes 
productions are suppressed, the tunneling wavefunction would not lead to 
catastrophic instability. Incidently, the Hartle-Hawking proposal also leads to 
the same mode functions for the scalar field as is obtained by Vilenkin. 
This implies the emergence of 
Bunch-Davies vacuum with the suppression of blackhole production rate 
substantiating the validity of Hartle-Hawking proposal and this is why Bousso     
and Hawking \cite{bou:prd} obtains the same conclusion from the wavefunction calculation.
In Bousso and Hawking, the blackhole production rate is given by 
\be
\Gamma = {{P_{\sc{S\, DS}}}\over {P_{\sc{DS}}}}= \exp{(-{1\over {3H^2}})},
\ee
where $P_{_{\sc{S\,DS}}}$ is the nucleation probability of a pair of blackholes in
$S^2 X S^1$ Schwarzchild-de Sitter universe and $P_{\sc{DS}}$ is the same in 
de Sitter universe without the pair of blackholes. The nucleation probability 
in Hartle-Hawking proposal is given by  
\be
P=\vert \psi_{_{\sc{H}}}\vert^2 \;\propto\; \exp{(-2\vert S_{_{\sc{E}}}\vert )},
\ee
at the nucleation point and (1) is obtained using the expression in (2) for 
$S\,DS$ and $DS$ background separately, $S_{_{\sc{E}}}$ being the euclidean action. 
It is argued that since the tunneling wavefunction $\psi_{_{\sc{T}}}$ grows like 
$e^{+\vert S_{_{\sc{E}}}\,\vert}$, 
hence (1) and (2) impart catastrophic instability for a tunneling 
wavefunction for small $H$.
\par  
In the present paper, we show that the wormhole dominance in the wavefunction 
and the concept of WKB Complex trajectory (CWKB) results in a most general 
wavefunction of the universe. The boundary conditions corresponding to the no 
boundary proposal and the tunneling proposal when introduced in 
$\psi_{_{\sc{WD}}}$ 
gives the respective wavefunctions. We also show that $\psi_{_{\sc{T}}}$ is also 
given by $\exp{(-S_{_{\sc{E}}})}$, when quantum corrections are taken into account. 
The quantum corrections is interpreted in our previous work \cite{bis:prd} as 
due to wormhole contribution. Hence the name is the wormhole dominance proposal.
Deviating from our previous work, we now interpret the quantum corrections in 
terms of Lorentzian sector, as well as in terms of wormhole contribution. If 
quantum corrections are not taken into the Hartle-Hawking wavefunction, it would 
have the drawback of not allowing sufficient inflation. Moreover we show that both 
the proposals give the same nucleation probability and hence we get the same 
Bunch-Davies vacuum in both the cases.
\section{\bf{Wormhole Dominance Proposal}}
We start with the Wheeler-DeWitt equation
\be
\left[ {{d^2}\over {da^2}}-a^2(1-H^2a^2)\,\right]\psi(a)=0,
\ee
for a homogeneous, isotropic and closed universe with constant vacuum energy 
$\rho_v$, where $H={4\over 3}G\rho_{v}^{1/2}$. The classical solution of the 
model is the de Sitter space
\be
a(t)=H^{-1}\cosh{(Ht)}.
\ee
The WKB solutions are :
\be
\bar{\psi^{\sc{\pm}}}=(a>H^{-1})=[p(a)]^{-1/2}\exp{\left[\,\pm i
\int^{a}_{H^{-1}}p(a')da'\mp i{\pi\over 4}\,\right]},
\ee 
\be
\tilde{\psi^{\sc{\pm}}}=(a<H^{-1})=[p(a)]^{-1/2}\exp{\left[\,\pm i
\int^{a}_{H^{-1}}p(a')da'\,\right]},
\ee 
where $p(a)=\left[\,-a^2(1-H^2a^2)\,\right]^{1/2}$. The wavefunctions 
corresponding to the Hartle-Hawking $(\equiv \psi_{_{\sc{H}}})$ and the tunneling 
$(\equiv \psi_{_{\sc{T}}})$ proposals are : 
\be
\psi_{_{\sc{H}}}(a<H^{-1})=
\tilde{\psi^-}\,(a),
\ee
\be
\psi_{_{\sc{T}}}(a<H^{-1})=\tilde{\psi^+}(a)-{i\over 2} \tilde{\psi^-}(a).
\ee
The CWKB solution of (3) is obtained as follows. Identifying 
\be
S(a_f,a_i)=\int_{a_i}^{a_f}p(a')da'\,,
\ee
where $a$ may be complex, the solution of (3) at a point $a$, real or complex 
is obtained as
\be
\psi(a)=\sum_{CWKB\;paths} \exp{\left[\,\pm iS(a,a_{\sc{0}})\,\right]}\,,
\ee
where $a_{\sc{0}}$ is an arbitrary point, where the boundary conditions are 
known or fixed. We now consider the classically unallowed region with 
$a<H^{-1}$. For a wave moving from left to right, we take the negative sign in 
(10) and 
call it the direct trajectory. The wave moving right to left is called 
reflected trajectory and corresponds to the positive sign. The classical 
trajectory corresponding to (3) is
\be
a(t)=H^{-1} \sin{(Ht)},
\ee
where $t$ may be complex. Here $a=0$ and $a=H^{-1}$ are the turning 
points and act as reflection point for trajectories that move towards it. 
We consider a point $a<H^{-1}$ 
and start from $a=0$. Thus in CWKB \cite{bis:prd}, neglecting the WKB preexponential 
factor
\beq
\psi (a)\sim (\psi_{_{\sc{DT}}}(a) -i\psi_{_{\sc{RT}}})X\,\;[\,Repeated\;\;
reflections\;\nonumber \\ 
between\;\;the\; \;turning\;\;points\;\;\,a=0,\;H^{-1}\,],
\eeq
where
\be
\psi_{_{\sc{DT}}}(a)\sim \exp{\left[ -iS(a,0)\,\right]},
\ee
\be
\psi_{_{\sc{RT}}}(a)\sim \exp{\left[ -iS(H^{-1},0)+iS(a,H^{-1})\,\right]},
\ee
\be
Repeated\;\;reflections\;\equiv{1\over {1-\exp{\left[\,-2iS(H^{-1},0)\,\right]}}}\,.
\ee
Using (12) to (15), we get for (3), using (9)
\be
\psi^\pm\,\sim 
{{\exp{({\pm{1\over {3H^2}}})}}
\over {1-\exp{({\pm{2\over {3H^2}}})}}}
\left[\,C_{_{\sc{\pm}}}\exp{(\mp {1\over {3H^2}}(1-a^2H^2)^{3/2})}
-d_{_{\sc{\pm}}}i\exp{(\pm{1\over {3H^2}}(1-a^2H^2)^{3/2})}\,\right]\,.
\ee
Here $C_{_{\sc{\pm}}}$ and $d_{_{\sc{\pm}}}$ are two constants, $\pm$ come from 
the negative sign under the square root in $p(a)$. Using (5,6), we write as
\be
\psi^\pm(a<H^{-1})=N_\pm\left[\, C_{_{\pm}}\tilde{\psi^\mp}(a)
-id_{_\pm}\tilde{\psi^\pm}(a)\,\right].
\ee
Eq.(17) is the most general wavefunction in CWKB. Here $N_\pm$ is given by the
prefactor outside the square bracket in (16). 
\par
Let us calculate the norm of the wavefunction $\psi(0_+)$ and nucleation 
probability according to (16) :
\be
\psi^+(0_+)={{-i(e^{{2\over {3H^2}}}+i)}\over
{1-e^{{2\over {3H^2}}}}}\,,
\ee
\be
\psi^-(0_+)
=-{{e^{{2\over {3H^2}}}-i}\over
{1-e^{{2\over {3H^2}}}}}\,,
\ee
so that we find $\vert\psi^+(0)\vert =\vert\psi^-(0)\vert$. This coincides with the norm 
given by Klebanov et al. \cite{kle:np} and which is identified due to 
wormholes contributions at $a=0$. The nucleation amplitude is given by
\be
\vert\psi^\pm(H^{-1})\vert
={{e^{\pm{2\over {3H^2}}}}\over
{1-e^{\pm{2\over {3H^2}}}}}\left[\,\vert C_{_\pm}\,\vert^2
+\vert d_{_\pm}\,\vert^2\,\right]^{1/2}\,.
\ee
Let us see how do the standard wavefunctions emerge from (17). In the present 
work we propose a more transparent interpretation of $N_\pm$ in terms of 
Lorentzian sector instead of wormhole dominance. In CWKB, $a(t)$ may be complex 
and we write  
\be
a(t)=H^{-1}\sinh{(t_{_{\sc{R}}}+it_{_{\sc{I}}})}\,.
\ee
Consider $t_{_{\sc{R}}}=H^{-1}{\pi\over 2}$, Eq. (21) gives
\be
a(t)\;\matrix{\longrightarrow \cr t_{_{\sc{R}}}\;=H^{-1}{\pi\over 2}}
H^{-1}\cosh{(Ht_{_{\sc{I}}})}.
\ee
This corresponds to a Lorentzian de Sitter universe and corresponds to outgoing 
and ingoing trajectories for $t_{_{\sc{I}}}>\,and\, t_{_{\sc{I}}} <0$. Thus a 
trajectory from $a=0$ to $a=H^{-1}$ and then parrallel to the imaginary axis 
gives rise to both the outgoing and ingoing modes where 
$t_{_{\sc{I}}}$ serves as Lorentzian time. In the 
region $a<H^{-1}$, the euclidean contribution is 
$\exp{\left[-iS(H^{-1},0)\,\right]}=\exp{({1\over {3H^2}})}$. As $H^{-1}$ is a 
turning point, the ingoing and outgoing modes must have equal amplitudes at the 
point $a=H^{-1}$. This corresponds to the Hartle-Hawking proposal. The 
wavefunction $\psi_{_{\sc{H}}}$ is then given by $\psi_{_{\sc{WD}}}^{+}$ with 
$C_+=1,\,d_+=0$ and hence no repeated reflections. Thus 
\be
\psi^{+}_{_{\sc{WD}}}(a<H^{-1})
=\exp{({1\over{3H^2}})}
\tilde{\psi^-}(a)
\equiv\psi_{_{\sc{H}}}(a<H^{-1}),
\ee
\be
\psi^{+}_{_{\sc{WD}}}(a>H^{-1})
=\exp{({1\over{3H^2}})}
({\bar{\psi^+}(a)-\bar{\psi^-}(a)})
\equiv\psi_{_{\sc{H}}}(a>H^{-1}).
\ee
The nucleation probability is
\be
\vert\psi_{_{\sc{H}}}(a=H^{-1})\vert^2=
\exp{({2\over {3H^2}})}.
\ee
When we consider $\psi^{-}_{_{\sc{WD}}}$, we should start from Lorentzian sector, 
where we have only outgoing modes from $H^{-1}$. This implies both growing and 
decaying exponentials and hence also repeated reflections between $a=0$ and 
$a=H^{-1}$. This corresponds to the tunneling proposal. Thus taking 
$\psi^{-}_{_{\sc{WD}}}(a<H^{-1})$ and $C_-=+1, d_-={1\over 2}$, we get
\be
\psi^{-}_{_{\sc{WD}}}(a<H^{-1})=
{{\exp{({-1\over {3H^2}})}}\over {1-{\exp{({-2\over {3H^2}})}}}}
\left[ \tilde{\psi^+}-{i/2} \tilde{\psi^-}(a)\,\right]
\equiv\psi_{_{\sc{T}}}(a<H^{-1}).
\ee
We have taken $d_-=1/2$ to have equal amplitude at the turning point. If we keep 
quantum corrections to both the proposals, we have
\be
\psi^{+}(a<H^{-1})=
{{\exp{({1\over {3H^2}})}}\over {1-{\exp{({2\over {3H^2}})}}}}
\tilde{\psi^-}(a),
\ee
\be
\psi^{-}(a<H^{-1})=
{{\exp{({-1\over {3H^2}})}}\over {1-{\exp{({-2\over {3H^2}})}}}}
\left[ \tilde{\psi^+}-{i/2} \tilde{\psi^-}\,\right],
\ee
and the probability of nucleation
\be
P(a=H^{-1})\equiv\vert \psi^{\sc{\pm}}(H^{-1})\vert^2=
{{\exp{({2\over {3H^2}})}}\over {({1-{\exp{({2\over {3H^2}})}}})^2}},
\ee
the same in the two proposals. This result is a bit surprising. Let us consider 
the most general expression of $\psi_{_{\sc{WD}}}^{\pm}$. For $a<H^{-1}$, the four 
sphere having the boundary as 3 sphere of radius $a$, the action is 
\be
S_\pm=-{1\over {3H^2}}\left[\,1\pm(1-H^2a^2)^{3/2}\,\right],
\ee
where the plus (minus) sign denotes the action that corresponds to filling in 
the 3 sphere with more (less) than half the 4 sphere. In terms of CWKB this 
corresponds to the reflected trajectory, and the direct trajectory. We now write 
the $\psi_{_{\sc{WD}}}^{\pm}$ in terms of $S_{\sc{\pm}}$. We get from (26)
\be
\psi^{+}(a<H^{-1})=
{1\over {1-{\exp{({2\over {3H^2}})}}}}
\left[\,C_+\,\exp{(-S_-)}-id_+\,\exp{(-S_+)}\,\right],
\ee
\be
\psi^{-}(a<H^{-1})=
-{1\over {1-{\exp{({2\over {3H^2}})}}}}
\left[\,C_-\,\exp{(-S_+)}-id_+\,\exp{(-S_-)}\,\right].
\ee
Thus we see that the quantum corrections arising out of repeated reflections do 
all the necessary job to cast the tunneling wavefunction also in the form 
$\psi\sim \exp{(-S_{_{\sc{E}}})}$. Not only that, the repeated reflections also save 
the Hartle-Hawking wavefunction from the drawback for not having sufficient 
inflation. If $H$ is small, then (27) gives     
\[\vert\psi^+(a=H^{-1})\vert^2\sim e^{-{2\over {3H^2}}},\]
with small nucleation probability for large universes.
\section{\bf{Conclusion}}
Thus our conclusion is that if we do not take quantum corrections either through 
wormhole dominance or repeated reflections at the turning points, the discord 
among the proposals would sustain. Our proposal in terms of CWKB paths gives a 
plausible answer to the current discord on the wavefunction of the universe. The
normalization and other aspects of $\psi_{_{\sc{WD}}}$ have already been 
discussed in our previous work \cite{bis:prd}. We have shown that 
$\psi_{_{\sc{T}}}$ now grows as $\exp{(-S_{_{\sc{E}}})}$ and hence Bousso and 
Hawking's criticism does not apply to it. 
\par
Allowance of repeated reflections in the tunneling proposal is quite natural 
since we have both $\psi^+$ and $\psi^-$ like terms in the region $a<H^{-1}$. 
But in the Hartle-Hawking proposal, it cannot be obtained since it has only 
$\tilde{\psi^-}$ like term in the regions $a<H^{-1}$. The wormholes require 
a contribution $\sim\exp{({2\over {3H^2}}-{1\over 2}a_{min}^2)}$ i.e., a 
$\tilde{\psi^+}$ -like term, where $a_{min}$ is the radius of the wormhole 
throat [see ref. \cite{bis:prd} and ref. \cite{kle:np}]. The absence of 
$(1-\exp{({2\over {3H^2}})})$ like term in $N_+$ would then imply for having not 
sufficient inflation in the Hartle-Hawking proposal.
\par
With respect to pair production of blackholes we mention that since we have the 
same nucleation probability in both the proposal, using Mottola's arguments 
\cite{mot:prd} 
we can show that energy density of the produced pairs is given by 
\be
T_{ab}\;\propto\;\vert\beta\vert g_{ab}\;\sim\; e^{-{{2\pi m}\over H}} g_{ab}.
\ee
Hence the suppression of blackholes for small $H$ and $m>>H$ is not forbidden by 
any of the proposals so long as $a=H^{-1}$ acts as reflection point. More details 
in this regard would be explored shortly. 

\smallskip
\section{\bf{Acknowledgment}}                          
\par
A.Shaw acknowledges the financial support from ICSC World Laboratory,
LAUSSANE during the course of the work.
\end{document}